\documentclass[seceq]{ptptex}
\newcommand{\beq}{\begin{equation}}
\newcommand{\eeq}{\end{equation}}
\newcommand{\beqa}{\begin{eqnarray}}
\newcommand{\eeqa}{\end{eqnarray}}
\newcommand{\beqas}{\begin{eqnarray*}}
\newcommand{\eeqas}{\end{eqnarray*}}
\newcommand{\bsym}{\boldsymbol}

\newcommand{\svec}[1]{\mbox{{\footnotesize \boldmath ${#1}$}}}
\newcommand{\gvec}[1]{\mbox{\boldmath ${#1}$}}

\title{%        %You can use \\ for explicit line-break.
Renormalization-group for amplitude equations in cellular pattern formation with and without conservation law%
}

\author{%       %Use \scshape for the family name.
Yasuhiro \textsc{SHIWA}\footnote{E-mail: shiway@kit.ac.jp}%
}

\inst{%     %Affiliation, neglected when [addenda] or [errata].
Statistical Mechanics Laboratory, Kyoto Institute of Technology, Matsugasaki, Sakyo-ku, Kyoto 606-8585, Japan
}

\recdate{\today} %Mmmmm DD, YYYY}%            %Editorial Office will fill in this.

\abst{%         %This abstract is neglected when [addenda] or [errata].
A proper version of the proto renormalization-group scheme is presented to derive amplitude equations in striped pattern formation with conserved and nonconserved order parameter. In the conserved case, the result preserves the conservation law as well as the rotational invariance of the physical system. This makes a great contrast with existing singular perturbation theories such as the multiple-scales method, in which the perturbation expansion truncated at any finite order destroys those symmetries.
}

\PTPindex{056, 054, 013}  %Input the subject index(es) of your paper, 
\begin{document}

\maketitle

\section{Introduction}

Spatial patterns are seen in a great variety of physico-chemical systems. They are typically periodic in space, at least locally. The most common are stripes (or often called rolls or lamellae) and hexagons. The unity of dynamical mechanisms of their birth and evolution allows us to consider some models of the nonlinear theory of such structures. The Swift-Hohenberg (SH) equation \cite{sh} is one of the simplest and most canonical paradigms, and has been intensively studied in the past.\cite{cross}

In order to describe the appearance and growth of crystalline phases that occurs in many technologically important materials processing, a dynamical model called phase-field crystal (PFC) model is also introduced.\cite{pfc} \ Since its introduction, the PFC model has been applied to a wide range of problems such as solidification, elastic and plastic deformations.\cite{overview_pfc} \ It is a continuum equation as the SH equation, and additionally there exists a global conservation law for the order parameter in this model. The presence of the conservation law is expected to give rise to new universal features of pattern formation which largely differ from those bestowed on the systems lacking the conserved quantity such as the SH model.\cite{csh} \ As a matter of fact, the PFC model is a conserving analogue of the SH model.

Since it is impossible to obtain analytical solutions to those models in pattern formation problems that are nonlinear partial differential equations, it is quite natural that a method to coarse-grain the model has been sought. Among others it is now relatively standard to use the so-called amplitude equation formalism. In the vicinity of the pattern forming instability, fluctuations of the order parameter field favor the formation of structures built on some finite wavevectors. The evolution of these unstable modes is much slower than the evolution of the stable ones. Hence the faster evolution of the stable modes may be adiabatically eliminated from the dynamics. As a result, the dynamics can be reduced to the slow dynamics of the amplitudes of these unstable modes. The amplitude equations that result have been a powerful tool in the study of pattern formation, selection and stability properties.\cite{walgraef} \ As such there exist many theoretical methods to derive them. Important examples include the method of multiple scales \cite{cross} and renormalization-group (RG) methods.\cite{chen,no} \ In particular, the proto-RG method which has been proposed in Ref.~\citen{no} is the most abstract version of the RG approach to free as much as possible the RG procedure from the necessity of obtaining explicit expressions of secular terms that are required of the other RG method.\cite{chen} \ It is therefore not surprising that the proto-RG scheme has been used to study the SH and PFC models.\cite{no,athreya} \ For the present account, it is to be noted in the latter paper that the authors conjectured that one would need separate RG schemes to coarse-grain order parameter equations depending on whether or not the order parameter is conserved. Yet the issue remains controversial,\cite{dispute} \ and in the present paper we have reformulated the proto-RG reduction of the SH and PFC models in a successful attempt to accommodate the dispute.

In the present paper we will thus concern ourselves with a system with a real order-parameter field $\psi$ whose dynamics are described by the zero-noise Langevin equation
\beq
\partial_{t}\psi=\left( i\partial_{\bsym{x}} \right)^{p}\left[ \epsilon\psi - \psi^{3}-(\partial_{\bsym{x}}^{2}+k_{0}^{2})^{2}\psi\right], \label{model_phi}
\eeq
where $p=0$ for a nonconserved order parameter (NCOP), $p=2$ for a conserved order parameter (COP). The $\partial_{\bsym{x}}$ is a two-dimensional gradient operator ($\nabla$) with respect to the position vector $\bsym{x}$, and $\partial_{\bsym{x}}^{2}$ is the Laplacian ($\nabla^{2}$). Hence the choice $p=0$ corresponds to the SH equation, while the PFC equation is represented by $p=2$. The $\epsilon$ is the bifurcation parameter. It is well known that the competition between the surface energy contribution given by $\partial_{\bsym{x}}^{2}\psi$ term in the square brackets on the right-hand side (RHS) of Eq. (\ref{model_phi}) and the curvature energy term $\propto \partial_{\bsym{x}}^{4}\psi$ gives rise to spatially modulated structures with period $\sim 2\pi/k_{0}$ for $\epsilon >0$. In the present discussion we restrict ourselves to striped patterns for the sake of brevity, although the generalization to hexagonal patterns is straightforward. In doing so, we demonstrate in the following \S 2 that by a suitable modification of the original scheme, \cite{no} proper amplitude equations can be derived on the same footing in the proto-RG formalism in both cases of NCOP and COP. As the essential logic of the formalism can be recognized at the lowest nontrivial order, in the present paper the explicit calculations will be exhibited to $O(\epsilon)$. We compare the analysis with the multiple-scales theory in \S 3, and concluding remarks are given in \S 4.
 
\section{The Proto-RG Scheme}

Since we treat the term $\epsilon \psi-\psi^{3}$ in Eq. (\ref{model_phi}) as a perturbative one, we scale $\psi$ as $\sqrt{\epsilon}\psi$ and denote the new $\psi$ with the same symbol. The model (\ref{model_phi}) then reads
\beq
\partial_{t}\psi=\left( i\partial_{\bsym{x}} \right)^{p}\left[ \epsilon(\psi - \psi^{3})-(\partial_{\bsym{x}}^{2}+k_{0}^{2})^{2}\psi\right]. \label{model}
\eeq
Throughout we take the average value of $\psi$ is zero.

We expand the solution to Eq.~(\ref{model}) as
\beq
\psi(t,\bsym{x})=\psi_{0}+\epsilon \psi_{1}+\epsilon^{2}\psi_{2}+\cdots.
\eeq
The $\psi_{j}, j=0,1,2,\cdots$, obey the equations
\beqa
L(\partial_{t},\partial_{\bsym{x}})\psi_{0}&&=0,\label{zeroth}\\
L(\partial_{t},\partial_{\bsym{x}})\psi_{1}&&=\left( i\partial_{\bsym{x}} \right)^{p}(\psi_{0}-\psi_{0}^{3}),\label{first}\\
L(\partial_{t},\partial_{\bsym{x}})\psi_{2}&&=\left( i\partial_{\bsym{x}} \right)^{p}\left\{ (1-3\psi_{0}^{2})\psi_{1}\right\},\label{second}\\\mbox{etc.},&& \nonumber
\eeqa
with
\beq
L(\partial_{t},\partial_{\bsym{x}})\equiv \partial_{t}+\left( i\partial_{\bsym{x}} \right)^{p}(\partial_{\bsym{x}}^{2}+k_{0}^{2})^{2}.\label{op_L}
\eeq
We have the zeroth-order solution from (\ref{zeroth}) in the form
\beq
\psi_{0}(\bsym{x},B,A)=B(\gvec{\rho})+\left( A(\gvec{\rho})\mbox{e}^{i\bsym{k}\cdot \bsym{x}}+c.c.\right),\ |\bsym{k}|=k_{0}, \label{psi0}
\eeq
where $B$ and $A$ are, respectively, the real and complex function of an arbitrary space parameter $\gvec{\rho}$, which will be later used as the regularizing point. Here and hereafter we must set $B=0$ for the NCOP case, i.e., for $p=0$. The dependence of the solution on $B, A$ is explicitly denoted for clarity of the subsequent discussion. Thus
\beq
\psi(t,\bsym{x})=B+\left( A\mbox{e}^{i\bsym{k}\cdot \bsym{x}}+c.c.\right)+\epsilon\psi_{1}(t,\bsym{x},B,A)+\epsilon^{2}\psi_{2}(t,\bsym{x},B,A)+\cdots. \label{psi}
\eeq
Notice that $\Phi_{0}\equiv \mbox{e}^{i\bsym{k}\cdot \bsym{x}}$ is the eigenfunction of the zero eigenvalue of $L$; $L\Phi_{0}=0$. It is also the case with $\Psi_{0}\equiv$ constant when $p=2$; henceforth we may arbitrarily set $\Psi_{0}=1$. Hence $\Phi_{0}$ and $\Psi_{0}$ can be the source of the secular terms in $\psi_{j}, j\ge 1$. 

Let $\hat{\psi}_{j}(r,R,B,A)$ be $\psi_{j}(r,B,A)$ with variables $r\equiv \{ t,\bsym{x}\}$ in the secular prefactors of $\Phi_{0}$ and $\Psi_{0}$ replaced by $R\equiv \{ \tau,\gvec{\rho}\}$, discarding the constant terms. The regularization of $\psi_{j}$ is made by introducing the renormalized $B$ and $A$, denoted by $B_{R}$ and $A_{R}$, via
\beqa
A&=&A_{R}Z\equiv A_{R}(R)\left[ 1+\epsilon Z_{1}(R)+\epsilon^{2}Z_{2}(R)+\cdots \right],\label{AR}\nonumber \\
B&=&B_{R}W\equiv B_{R}(R)\left[ 1+\epsilon W_{1}(R)+\epsilon^{2}W_{2}(R)+\cdots \right].\label{BR}
\eeqa
Then Eq.~(\ref{psi}) can be written as 
\beqa
\psi&=&B_{R}\Psi_{0}+(A_{R}\Phi_{0}+c.c.)+\epsilon\Bigl[ B_{R}W_{1}\Psi_{0}+(A_{R}Z_{1}\Phi_{0}+c.c.)+\psi_{1}(r,B_{R},A_{R})\Bigr]+\nonumber\\
  & & \epsilon^{2}\Bigl[ B_{R}W_{2}\Psi_{0}+(A_{R}Z_{2}\Phi_{0}+c.c.)+\psi_{2}(r,B_{R},A_{R})+\Bigr. \nonumber\\
  & & \Bigl. \hspace*{18pt}B_{R}W_{1}\partial_{B}\psi_{1}(r,B_{R},A_{R})+\bigl( A_{R}Z_{1}\partial_{A}\psi_{1}(r,B_{R},A_{R})+c.c.\bigr)\Bigr] +\cdots.\label{psi2}
\eeqa
The renormalization constants $Z_{j}$ and $W_{j}$ are determined order by order via
\beqa
&& B_{R}W_{1}\Psi_{0}+\left( A_{R}Z_{1}\Phi_{0}+c.c.\right)+\hat{\psi}_{1}(r,R,C_{R})=0,\nonumber\\
&& B_{R}W_{2}\Psi_{0}+\left( A_{R}Z_{2}\Phi_{0}+c.c.\right)+\hat{\psi}_{2}(r,R,C_{R})+\nonumber\\
&& \hspace*{36pt} B_{R}W_{1}\partial_{B}\hat{\psi}_{1}(r,R,C_{R})+\left(A_{R}Z_{1}\partial_{A}\hat{\psi}_{1}(r,R,C_{R})+c.c.\right)=0,\nonumber\\
&& \cdots,\label{Z}
\eeqa
where $C_{R}\equiv\{ B_{R}, A_{R} \}$.
It is then easy to see that Eqs.~(\ref{Z}) and (\ref{psi2}) are obtained by the $\epsilon-$expansion of the following equations, respectively:
\beq
(B-B_{R})\Psi_{0}+\bigl( (A-A_{R})\Phi_{0}+c.c.\bigr)+\epsilon\hat{\psi}_{1}(r,R,C)+\epsilon^{2}\hat{\psi}_{2}(r,R,C)+\cdots=0,\label{a_rg}
\eeq
\beqa
\psi&=&B_{R}\Psi_{0}+\left( A_{R}\Phi_{0}+c.c. \right)+\nonumber\\
& & \epsilon\left[ \psi_{1}(r,C)-\hat{\psi}_{1}(r,R,C)\right]+\epsilon^{2}\left[ \psi_{2}(r,C)-\hat{\psi}_{2}(r,R,C)\right]+\cdots.\label{psi_rg}
\eeqa
In the above $C\equiv \{B,A\}$. Notice that all the secular terms in $\psi_{j}$'s are removed in Eq. (\ref{psi_rg}).

Introduction of the variables $R$ is equivalent to splitting the derivative $\partial_{t}$ to $\partial_{t}+\partial_{\tau}$, and $\partial_{\bsym{x}}$ to $\partial_{\bsym{x}}+\partial_{\svec{\rho}}$. Since $\psi$ (and $\psi_{j}$) must be independent of the arbitrary regularization point $R$, we have 
${\cal F}\psi=0$ and ${\cal F}\psi_{j}=0$, where ${\cal F}\equiv  L(\partial_{t}+\partial_{\tau},\partial_{\bsym{x}}+\partial_{\svec{\rho}})-L(\partial_{t},\partial_{\bsym{x}})$. Applying the operator ${\cal F}$ to Eq.~(\ref{psi_rg}), we then find
\beq
L(\partial_{t}+\partial_{\tau},\partial_{\bsym{x}}+\partial_{\svec{\rho}})\bigl[ B_{R}\Psi_{0}+(A_{R}\Phi_{0}+c.c.)\bigr]=\epsilon{\cal F}\hat{\psi}_{1}(r,R,C)+\epsilon^{2}{\cal F}\hat{\psi}_{2}(r,R,C)+\cdots.\label{LA}
\eeq
Let ${\cal P}$ be the projection onto $\Phi_{0}$. Then by the definition of $\hat{\psi}_{j}$ we have
\beq
{\cal P}L(\partial_{t},\partial_{\bsym{x}})\hat{\psi}_{j}(r,R,C)=0.\label{id1}
\eeq
Hence
\beqa
{\cal P}L(\partial_{t}+\partial_{\tau},\partial_{\bsym{x}}+\partial_{\svec{\rho}})A_{R}\Phi_{0}&=&\epsilon{\cal P}L(\partial_{t}+\partial_{\tau},\partial_{\bsym{x}}+\partial_{\svec{\rho}})\hat{\psi}_{1}(r,R,C)+\nonumber\\
   & & \epsilon^{2}{\cal P}L(\partial_{t}+\partial_{\tau},\partial_{\bsym{x}}+\partial_{\svec{\rho}})\hat{\psi}_{2}(r,R,C)+\cdots. \label{protoRGeqn_A}
\eeqa
Similarly, using the projection operator onto $\Psi_{0}$, ${\cal Q}$, we have
\beqa
{\cal Q}L(\partial_{t}+\partial_{\tau},\partial_{\bsym{x}}+\partial_{\svec{\rho}})B_{R}\Psi_{0}&=&\epsilon{\cal Q}L(\partial_{t}+\partial_{\tau},\partial_{\bsym{x}}+\partial_{\svec{\rho}})\hat{\psi}_{1}(r,R,C)+\nonumber\\
   & & \epsilon^{2}{\cal Q}L(\partial_{t}+\partial_{\tau},\partial_{\bsym{x}}+\partial_{\svec{\rho}})\hat{\psi}_{2}(r,R,C)+\cdots. \label{protoRGeqn_B}
\eeqa
Equations (\ref{protoRGeqn_A}) and (\ref{protoRGeqn_B}) are the basic equations of the proto-RG scheme. We remark that the argument of $\hat{\psi}_{j}$ is $C$, not $C_{R}$. Thus, in the proto-RG formulation, there is no such ordering ambiguity as suggested in Ref.~\citen{athreya} for renormalization and differential operations. Consequently, there should not be a different ordering (renormalization before differentiation) as the authors proposed for the COP equation.

We shall now carry out the $O(\epsilon)$ computation of Eq. (\ref{protoRGeqn_A}). If we apply $L(\partial_{t}+\partial_{\tau},\partial_{\bsym{x}}+\partial_{\svec{\rho}})$ to $\hat{\psi}_{1}$ and separate out terms containing $\Phi_{0}$ from the outcome, then from Eq.~(\ref{first}) it must be identical to the coefficient of $\Phi_{0}$ in 
\beq
\bigl[ i (\partial_{\bsym{x}}+\partial_{\svec{\rho}})\bigr]^{p}(\psi_{0}-\psi_{0}^{3}). 
\eeq
Hence, explicitly written, Eq.~(\ref{protoRGeqn_A}) reads
\beqa
\lefteqn{{\cal P}\Bigl\{ \partial_{t}+\partial_{\tau}+\bigl[ i (\partial_{\bsym{x}}+\partial_{\svec{\rho}})\bigr]^{p}\left[ (\partial_{\bsym{x}}+\partial_{\svec{\rho}})^{2}+k_{0}^{2}\right]^{2}\Bigr\} A_{R}\Phi_{0}}\nonumber\hspace{6cm}\\
       &=& \epsilon{\cal P}\bigl[ i (\partial_{\bsym{x}}+\partial_{\svec{\rho}})\bigr]^{p}(\psi_{0}-\psi_{0}^{3})\label{protoRGeqn_A2}
\eeqa
to $O(\epsilon)$. 

At this juncture, we note 
\beqa
\psi_{0}^{3}-\psi_{0}&=& B^{3}+6B|A|^{2}-B +\nonumber\\
    & & \left\{\mbox{e}^{i\bsym{k}\cdot\bsym{x}}(3B^{2}A+3|A|^{2}A-A)+\mbox{e}^{2i\bsym{k}\cdot\bsym{x}}3BA^{2}+\mbox{e}^{3i\bsym{k}\cdot\bsym{x}}A^{3}+c.c.\right\}.
\eeqa
This, together with the identity: 
\beq
(\partial_{\bsym{x}}+\partial_{\svec{\rho}})^{2}f(\gvec{\rho})\Phi_{0}=\Phi_{0}({\cal L}_{k}-k_{0}^{2})f(\gvec{\rho}), \ \ {\cal L}_{k}\equiv \partial_{\svec{\rho}}^{2}+2i\bsym{k}\cdot \gvec{\partial}_{\svec{\rho}}
\eeq
for an arbitrary function of $\gvec{\rho}$, $f(\gvec{\rho})$, enable us to write down Eq.~(\ref{protoRGeqn_A2}) explicitly. That is,
\beq
\partial_{\tau}A_{R}+{\cal L}_{k}^{2}A_{R}=\epsilon(A-3|A|^{2}A)\ \  \mbox{(NCOP)}\label{protoRG_ncop}
\eeq
for $p=0$, and
\beq
\partial_{\tau}A_{R}-({\cal L}_{k}-k_{0}^{2}){\cal L}_{k}^{2}A_{R}=\epsilon({\cal L}_{k}-k_{0}^{2})(3B^{2}A+3|A|^{2}A-A)\ \ \mbox{(COP)}\label{protoRG_A_cop}
\eeq
for $p=2$. 

To calculate Eq. (\ref{protoRGeqn_B}) in the case $p=2$, we may follow much the same procedure as that used above to derive Eqs. (\ref{protoRG_ncop}) and (\ref{protoRG_A_cop}) from Eq. (\ref{protoRGeqn_A}). The $O(\epsilon)$ result is
\beq
\partial_{\tau}B_{R}-(\partial_{\gvec{\rho}}^{2}+k_{0}^{2})^{2}\partial_{\gvec{\rho}}^{2}B_{R}=\epsilon\partial_{\gvec{\rho}}^{2}(B^{3}+6|A|^{2}B-B).\label{protoRG_B_cop}
\eeq
 
Here we use $A=A_{R}Z, B=B_{R}W$ on the RHS in Eqs. (\ref{protoRG_ncop})--(\ref{protoRG_B_cop}), and then we may set $\gvec{\rho}=0$ and hence $Z=W=1$ since the RHS should not depend on $\gvec{\rho}$ explicitly. Therefore we finally obtain the proto-RG equations. Replacing the now dummy variables as $A_{R} \rightarrow A, B_{R} \rightarrow B, \gvec{\rho} \rightarrow \bsym{x}, \tau \rightarrow t$, they read
\beq
\partial_{t}A=(\epsilon - \Box^{2})A-3\epsilon|A|^{2}A\ \ \ (NCOP),\label{A_ncop}
\eeq
and 
\beqa
\partial_{t}A&=&(\Box - k_{0}^{2})\bigl[ (\Box^{2}-\epsilon)A+3\epsilon(|A|^{2}+B^{2})A\bigr], \label{A_cop}\\
\partial_{t}B&=&\nabla^{2}\bigl[ (\nabla^{2}+k_{0}^{2})^{2}B+\epsilon(B^{3}-B)+6\epsilon|A|^{2}B\bigr]\ \ \ (COP),\label{B_cop}
\eeqa
where $\Box$ is the rotationally covariant operator
\beq
\Box\equiv \nabla^{2}+2i\bsym{k}\cdot \nabla. \label{covariantderiv}
\eeq
The result (\ref{A_ncop}) is exactly of the same form as the amplitude equation for the SH model that Gunaratne {\it et al.} \cite{gunaratne} obtained first with a use of the multiple-scales method. Naturally, it also agrees with the proto-RG equation that Oono {\it et al.} \cite{no} derived when they first formulated the proto-RG scheme. A set of proto-RG equations (\ref{A_cop}) and (\ref{B_cop}) is new. As a matter of fact, Eqs. (\ref{A_cop}) and (\ref{B_cop}) are identical to the results given in Ref.~\citen{ms_pfc} which are obtained with the multiple-scales method truncated at $O(\epsilon^{7/2}$); see, however, the caveat discussed in the next section. 

\section{Discussion}

Notice that the appearance of the operator $\Box-k_{0}^{2}$ in Eq.~(\ref{A_cop}) can be directly traced back to the use of the operator $(\partial_{\bsym{k}}+\partial_{\gvec{\rho}})^{2}$ in the formula (\ref{protoRGeqn_A}). In fact, each of the higher-order contributions from the RHS of this formula adds to the RHS of Eq.~(\ref{A_cop}) the terms of the form $\epsilon^{n}(\Box-k_{0}^{2})\times$ function of $A,B$, $n=2,3,\cdots$. Thus Eq. (\ref{protoRGeqn_A}) in the case of COP yields the proto-RG equation of the form
\beq
\partial_{t}A=(\Box-k_{0}^{2})\left[ (\Box^{2}-\epsilon )A+3\epsilon(|A|^{2}+B^{2})A+h.o.t.\right],\label{protoRG_A2}
\eeq
where $h.o.t.$ denotes the higher-order (in $\epsilon$) terms. Multiplying by $\Phi_{0}$ and integrating the above equation (\ref{protoRG_A2}), we find 
\beq
\frac{d}{dt}\int d\bsym{x}A(t, \bsym{x})\Phi_{0}(\bsym{x})=0 \label{conA}
\eeq
owing to the presence of the operator $\Box-k_{0}^{2}$. The similar line of argument can be applied to Eq. (\ref{protoRGeqn_B}). Thus we find the proto-RG equation for $B$ takes the form $\partial_{t}B=\nabla^{2}[\cdots]$, which then guarantees that
\beq
\frac{d}{dt}\int d\bsym{x}B(t, \bsym{x})=0. \label{conB}
\eeq
Taken altogether, Eqs. (\ref{conA}) and (\ref{conB}) are in accord with the order-parameter conservation, $\int d\bsym{x}\psi(t, \bsym{x})=$constant, inherent in the COP equation. This in turn assures that the proto-RG equation respects the conservation law (and the rotational invariance as well) irrespective of at which order in $\epsilon$ the perturbation expansion in the RHS of Eqs. (\ref{protoRGeqn_A}) and (\ref{protoRGeqn_B}) is truncated.

On the other hand, it is not the case with the multiple-scales formalism. The fact that in this formalism the conserving operator $\Box-k_{0}^{2}$ is of mixed-order in $\epsilon$ forces the very important conclusion that the finite-order result always breaks the conservation law.\footnote{Incidentally, the importance of realizing that the rotationally covariant operator $\Box$ is of mixed-order in $\epsilon$ in the multiple-scales analysis for the SH model is emphasized in Ref.~\citen{gunaratne}.} \ To see this, let us recall the basic strategy of the so-called multiple-scales analysis; the reader should be warned that in fact it is not a multiple-scales (MS) analysis alone that is used but the combination of the MS analysis and reconstitution (hence called an MSR method). In the MSR analysis, first we carry out the MS analysis. Namely, we expand $\psi$ as a power series in the parameter $\delta$ where $\epsilon=\delta^{2}$;
\beq
\psi=\psi_{0}+ \delta\psi_{1}+\delta^{2}\psi_{2}+\cdots,
\eeq
and introduce slow space and time variables:
\beq
\bsym{X}=\delta \bsym{x},\ \ \ T=\delta^{2} t .\label{scales}
\eeq
The solution can be written as
\beqa
\psi&=& \Bigl( A_{10}(T,\bsym{X})+\delta A_{20}(T,\bsym{X})+\cdots \Bigr) +\nonumber\\
  & & \Big\{ \bigl( A_{11}(T,\bsym{X})+\delta A_{21}(T,\bsym{X})+\cdots \bigr) \mbox{ e}^{i\bsym{k}\cdot \bsym{x}} + \cdots + c.c \Big\},
\eeqa
where the last ellipsis represents terms whose wavenumber differs from critical, being generated by the nonlinear interactions between the basic modes. At respective orders in $\delta$, the Fredholm alternative imposes the condition on the dynamics of $A_{11}, A_{10}, \cdots$. We now employ a kind of resummation procedure to combine the obtained conditions. To that end we introduce the reconstituted amplitude functions for the critical mode:
\beq
A\equiv A_{11}+\delta A_{21}+\cdots,
\eeq
as well as for the zero mode: $B\equiv A_{10}+\delta A_{20}+\cdots$. We can then deduce the amplitude equations for the dynamics of $A$ and $B$. The operator (\ref{covariantderiv}) is the spatial derivatives acting on the envelope function $A$ and operates in the combination 
\[ 2i\delta \bsym{k}\cdot \nabla_{\bsym{X}}+\delta^{2}\nabla_{\bsym{X}}^{2} \]
where $\nabla_{\bsym{X}}$ operates on the slow variables $\bsym{X}$. When one generates the amplitude equation through the MSR method and if its perturbative expansion is carried out to all orders, then the amplitude equation will certainly guarantee the conservation law of the order-parameter field. However, this is not the case for finite truncation of the perturbation. For example, the correct term proportional to 
\[ \epsilon^{2}(\Box-k_{0}^{2})A|A|^{4} \]
will appear on the RHS of Eq.~(\ref{protoRG_A2}) if the amplitude equation is truncated at $O(\delta^{6})$, whereas at $O(\delta^{4})$ we only obtain $-\epsilon^{2}k_{0}^{2}A|A|^{4}$ and with this term present the conservation is lost. To state it differently, the symmetry-breaking term $-\epsilon^{2}k_{0}^{2}A|A|^{4}$ appearing at lower order in the perturbation expansion is symmetrized to restore the conservative symmetry only when the expansion is carried out to higher order. Since at any finite order there always appear such non-conserving terms, we cannot escape the afore-mentioned conclusion. Thus we see that, in the MSR analysis, in order to get the amplitude equation equivalent to our RG form (\ref{A_cop}) one needs to ``enforce" by hand the required conservation, which is obviously not a systematic step in the perturbation theory and is an uncontrollable approximation. In this connection one must always keep in mind that the MSR method is sometimes unreliable, and we refer the reader to Ref.~\citen{knobloch} for the danger of the reconstitution.

\section{Summary and Remarks}

We have presented a proper version of the proto-RG scheme to derive amplitude equations from NCOP and COP equations that produce spatially periodic patterns. The coarse-graining of the order parameter equations was performed by the proto-RG method. Our method is generically implemented the way in which the conservation law is preserved in the case of COP. There are two important aspects to be noted in our present formulation.

The first is that `integral constants' $B$ and $A$ in the solution (\ref{psi0}) depend on the renormalization point $\gvec{\rho}$. This lends a contrast to the conventional proto-RG formalism.\cite{no} \ The underlying idea is akin to the one imparted in the geometrical interpretation of the RG method \cite{envelope} in terms of the theory of envelopes. It was shown there (we are paraphrasing here in terms of envelope curves) that the RG method is a theory manipulating the `boundary conditions' at $x=\rho$. That is, suppose we have a perturbative solution which is valid only locally around $x=\rho$. Taking then $\rho$ as a running parameter, we have a family of curves (i.e., perturbative solutions), each curve being a good approximation around $x=\rho$. Then it turns out that the envelope which contacts with each local curve at the point of tangency $x=\rho$ gives a renormalized (global) solution.

Secondly, Eqs. (\ref{A_ncop})--(\ref{B_cop}) are the $O(\epsilon)$ formal calculation result of the proto-RG formulation. The order of the operator $\Box-k_{0}^{2}$ or $\Box$ is not yet determined in this method. In order to complete the RG procedure, we must carry out the reduction further by introducing the multiple scales like (\ref{scales}). The resulting equation is called the RG equation.\cite{no} \ Basically all that is required of such a reduction from proto-RG to RG equation is to find out the appropriate $\epsilon$-dependence of space-time scales, or what phenomenon you wish to see, which makes the RG equation consistent. In this reduction, there is always a risk of losing the relevant symmetry (and/or conservation law) for the reason expounded in \S 3. Your wish may or may not be granted. If you find the resulting RG equation is not consistent, so be it; it simply says your way of looking at the system is not consistent with the symmetry or conservation to the order you wish. For example, if you wish to require conservation to all orders (rigorous conservation) in the case of the proto-RG Eq. (\ref{A_cop}), with the choice of scales $\nabla \sim \epsilon^{1/4} \sim \bsym{k}$ it is consistent with $\partial_{t} \sim \epsilon^{3/2}$. In this case, we have proto-RG eq. = RG eq., and consequently our RG equation preserves the symmetry and conservation law at the nontrivial lowest order. Finally, we remark that there is no generic reason why the RG equation should agree with an MSR result. It is particularly the case if the MSR result in hand is obtained with an inappropriate choice of the scales. A more detailed account and application of these ideas, however, is beyond the scope of the present paper, and it will be published in a separate paper.

\end{document}